\documentclass[
aps,
superscriptaddress,
11pt,
endfloats,
]{revtex4-1}

\usepackage{graphicx, color, textcomp, stmaryrd}
\usepackage{wrapfig}
\usepackage{bm}
\usepackage{amsmath}              
\usepackage{amssymb} 
\usepackage[ngerman,english]{babel}
\usepackage{hyperref}
\usepackage{multirow}
\usepackage{placeins}

\begin{document}

\preprint{AIP/123-QED}

\title{Strain induced superconductivity in the parent compound BaFe\texorpdfstring{$_2$}{2}As\texorpdfstring{$_2$}{2}}

\author{J.~Engelmann}
\email{j.engelmann@ifw-dresden.de} 
\affiliation{IFW Dresden, Helmholtzstr.\ 20, 01069 Dresden, Germany}
\affiliation{TU Dresden, 01062 Dresden, Germany}
\author{V.~Grinenko} 
\email{v.grinenko@ifw-dresden.de}
\affiliation{IFW Dresden, Helmholtzstr.\ 20, 01069 Dresden, Germany}
\author{P.~Chekhonin} 
\author{W.~Skrotzki}
\affiliation{TU Dresden, 01062 Dresden, Germany}
\author{D.~V.~Efremov} 
\author{S.~Oswald} 
\author{K.~Iida} 
\author{R. H\"uhne} 
\author{J.~H\"anisch} 
\affiliation{IFW Dresden, Helmholtzstr.\ 20, 01069 Dresden, Germany}
\author{M.~Hoffmann}
\author{F.~Kurth}
\author{L.~Schultz}
\affiliation{IFW Dresden, Helmholtzstr.\ 20, 01069 Dresden, Germany}
\affiliation{TU Dresden, 01062 Dresden, Germany}
\author{B.~Holzapfel}
\affiliation{IFW Dresden, Helmholtzstr.\ 20, 01069 Dresden, Germany}
\affiliation{Karlsruhe Institute of Technology (KIT), Hermann-von-Helmholtz-Platz 1, 76344 Eggenstein-Leopoldshafen, Germany}

\begin{abstract}

The discovery of superconductivity (SC) with a transition temperature, $T_{\mathrm{c}}$, up to 65\,K in single-layer FeSe (bulk $T_{\mathrm{c}}$\,=\,8\,K) films grown on SrTiO$_3$ substrates has attracted special attention to Fe-based thin films. The high $T_{\mathrm{c}}$ is a consequence of the combined effect of electron transfer from the oxygen-vacant substrate to the FeSe thin film and lattice tensile strain. Here we demonstrate the realization of SC in the parent compound BaFe$_2$As$_2$ (no bulk $T_{\mathrm{c}}$) just by tensile lattice strain without charge doping. We investigate the interplay between strain and SC in epitaxial BaFe$_2$As$_2$ thin films on Fe-buffered MgAl$_2$O$_4$ single crystalline substrates. The strong interfacial bonding between Fe and the FeAs sublattice increases the Fe-Fe distance due to the lattice misfit which leads to a suppression of the antiferromagnetic spin density wave and induces SC with bulk-$T_{\mathrm{c}}$ $\approx$\,10\,K. 
These results highlight the role of structural changes in controlling the phase diagram of Fe-based superconductors.
 
\end{abstract} 

\maketitle

\section*{Introduction}

The discovery of Fe-based superconductors (FeSC) with critical temperatures, $T_{\mathrm{c}}$, exceeding 50\,K has
stimulated many experimental and theoretical efforts to reveal the mechanism for
superconductivity (SC) and to find compounds with even higher $T_{\mathrm{c}}$. \cite{Kamihara2008,Stewart2011,Chubukov2012}
The band structure of the  stoichiometric parent compounds yields two  or three  small
hole pockets around $\Gamma=(0,0)$ and two  electron pockets around $M=(\pi,\pi)$ in
the folded zone.\cite{Yaresko2009,Kordyuk2012}  The nesting between electron and hole bands makes the commensurate spin density wave structure (SDW) with $\textbf{Q}=(\pi,\pi)$ plausible. 
In all FeSCs the transition from the paramagnetic to the magnetic ordered SDW phase is accompanied by the structural transition from the tetragonal $I4/mmm$ to the orthorhombic $Fmmm$ phase. 
Doping due to injection of charge carriers (electrons or holes), chemical pressure and related disorder impairs the nesting conditions resulting 
in reduced structural transition temperature, $T_{\mathrm{s}}$, 
and N\'{e}el temperature, $T_{\mathrm{N}}$. At some finite doping the magnetic phase gives room to SC. The vicinity to the SDW phase suggests remaining strong spin fluctuations are possibly responsible for the high temperature SC. 
Therefore, the influence of doping on the phase diagram of FeSC is rather complex. 

Another way to realize SC in FeSC is applying external pressure.\cite{Kimber2009,Yamazaki2010} Finite pressure suppresses the SDW and restores the tetragonal symmetry. 
As a result, SC appears with $T_{\mathrm{c}}$'s comparable to that obtained by doping. 
However, the existence of SC, the value of maximum $T_{\mathrm{c}}$, and
the pressure under which SC occurs can vary considerably, 
depending on the level of hydrostaticity. \cite {Sefat2011}
By investigating the pressure media with different levels of hydrostaticity, 
it was found that a uniaxial component of the pressure dramatically affects the phase diagram of AEFe$_2$As$_2$ (AE122) (AE\,=\,Alkali earth Element) 
compounds.\cite {Yamazaki2010,Duncan2010,Sefat2011} It was suggested 
that the uniaxial pressure component is especially efficient for suppressing SDW order and responsible for a high $T_{\mathrm{c}}$ of $\approx$ 35\,K.
\cite {Yamazaki2010,Duncan2010,Sefat2011} 
However, to the best of our knowledge, SC under uniaxial pressure has been only reported in CaFe$_2$As$_2$ with a relatively low $T_{\mathrm{c}}$ about 10\,K.\cite{Torikachvili2009}  
Therefore, further experimental investigations are required to clarify the relevance of uniaxial pressure for stabilizing the SC state in FeSC.

In this article we demonstrate SC in stoichiometric parent compound Ba122 thin films solely by strain. 
We focus on epitaxial FeSC thin films, which are currently of great 
interest due to the ability to modify their electronic properties solely by 
change of the crystal structure in a well defined way.
\cite{Huang2010} Therefore, this method could aid understanding the conditions which can increase $T_{\mathrm{c}}$ of FeSC. 
Recently it was demonstrated that the bonding distances in FeAs planes might be altered by the 
lattice misfit to the substrate in Co-doped BaFe$_2$As$_2$ (Ba122)~\cite{Iida2009,kurth2013} and in 
FeSe$_{0.5}$Te$_{0.5}$~\cite{Bellingeri2011}. As a result, variation of 
$T_{\mathrm{c}}$ as a function of bonding distances was observed. In particular, it was discovered that SC with a transition temperature of 65\,K in single-layer FeSe 
\cite{He2013} can be induced by charge doping from the oxygen-vacant substrate to the thin film \cite{Tan2013} (bulk $T_{\mathrm{c}}$\,=\,8\,K \cite{Hsu2008}). However, 
it was emphasized that the effect of strain also plays an important role in 
driving SC.\cite{Tan2013} In order to distinguish between the effects of charge doping and strain on the phase diagram of FeSC, further investigations are necessary where these two effects are decoupled. Also, until now all the measurements regarding thin films were done for Fe-pnictide compounds which already demonstrate SC in the bulk form. 
Additionally, we show that our data provide a deeper understanding of the mechanism responsible for controlling the phase diagram 
of isovalently Ru-doped Ba122. 

\section*{Results}

\subsection*{Structural analysis} 

\begin{figure*}
	\centering
		\includegraphics[width=1.00\textwidth]{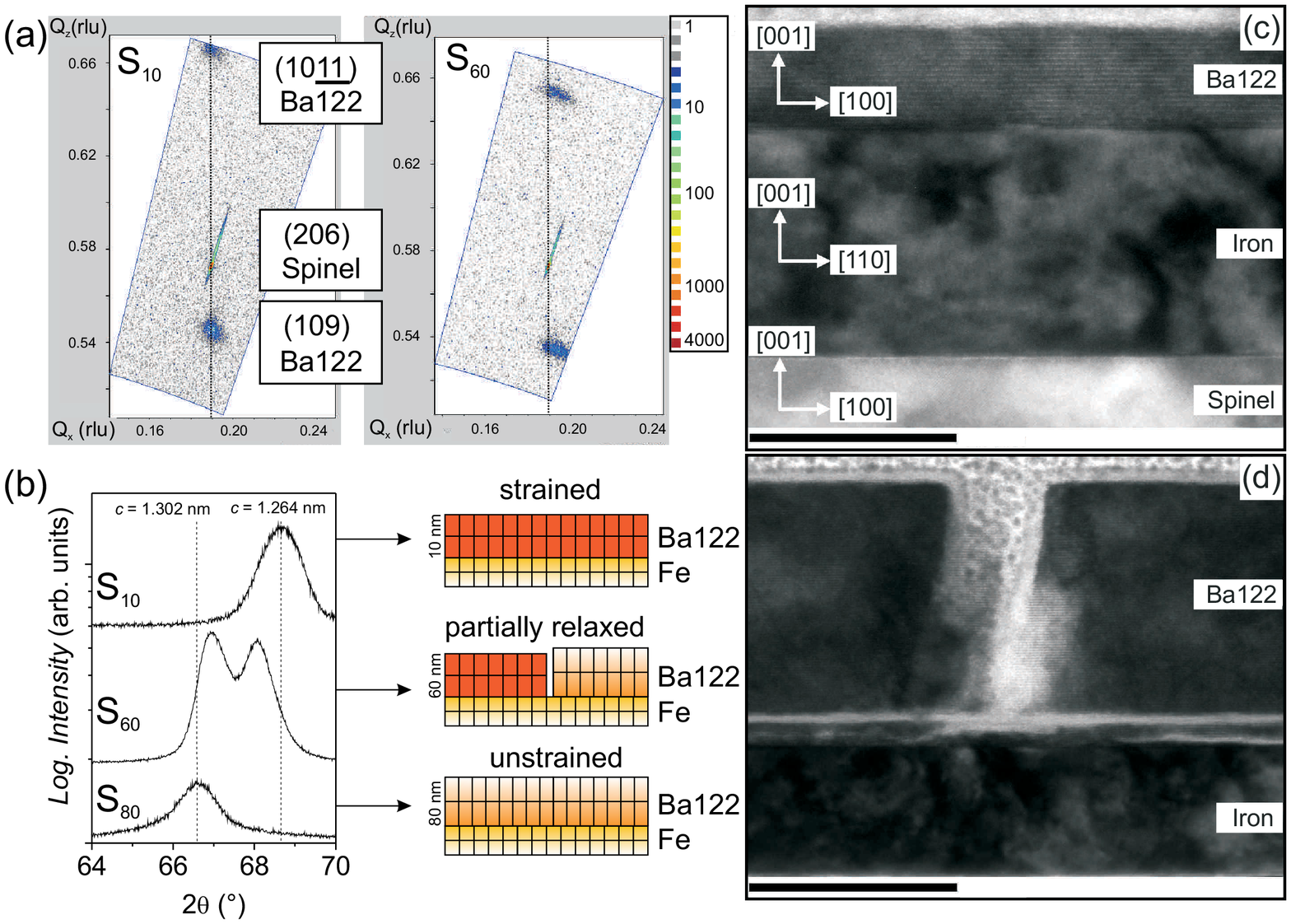}
		\caption{\textbf{Structural characterization by x-ray diffraction and transmission electron microscopy.} 
		(a) Reciprocal space maps of samples S$_{10}$ and S$_{60}$. Relaxation processes for sample S$_{60}$ are clearly seen as a 
		shift of the Ba122 reflections with regard to the substrate peak. The line is a guide for the eyes.  
	(b) Evolution of the (008) diffraction peak for sample S$_{10}$, S$_{60}$, and S$_{80}$ with corresponding sketches of 
	the strain state of the Ba122 layer. 
	The epitaxial growth of all Ba122 films was confirmed by $\theta$-2$\theta$ 
	x-ray diffraction in Bragg Brentano geometry and pole figure measurements (see Supplementary Figure S1). 
	Bright field TEM micrographs of two thin films: (c) sample S$_{10}$ (scale bar 25 nm) and (d) sample S$_{60}$ (scale bar 50 nm).}
	\label{figure1}
\end{figure*}

High resolution reciprocal space maps (RSM) (fig. \ref{figure1}a) of 
the Ba122 (1\,0\,9) and (1\,0\,\underline{11}) 
diffraction peaks as well as the (2\,0\,6) peak 
of the MgAl$_2$O$_4$ (spinel) substrate reveal that substrate and thin 
Ba122 layers ($d\,\lesssim\,d_{\mathrm{c}}$, where $d_{\mathrm{c}}$ is the critical thickness of Ba122) exhibit the same $a$ lattice constant ($a$\,=\,0.404\,nm). 
Increasing the thickness of the Ba122 layer $d$\,$>$\,$d_{\mathrm{c}}$ results 
in a relaxation of this layer, indicating 
a decrease of the $a$ lattice parameter (visualized by a shift of the 
Ba122 diffraction peaks to larger $Q_{\mathrm{x}}$ values). 
In the $\theta$-2$\theta$ scans (fig. \ref{figure1}b) a clear trend is 
observed with increasing thickness, $d$, of Ba122. 
The film with $d$\,=\,10\,nm (S$_{10}$) has a $c$ lattice parameter of 1.267\,nm, which is reduced by about 3\,\% in comparison to the target value ($c$\,=\,1.301\,nm). For $d$\,=\,60\,nm two different $c$ lattice parameters were found 
($c_1$\,=\,1.279~nm and $c_2$\,=\,1.301\,nm) which is explained by 
partial stress relaxation within the Ba122 volume. 
Further increasing $d$ results in complete stress relaxation within the whole 
volume of the Ba122 film. 
The whole set of lattice constants depending on the film thickness is summarized in table \ref{paras}.

\begin{figure}
	\centering
		\includegraphics[width=1.00\textwidth]{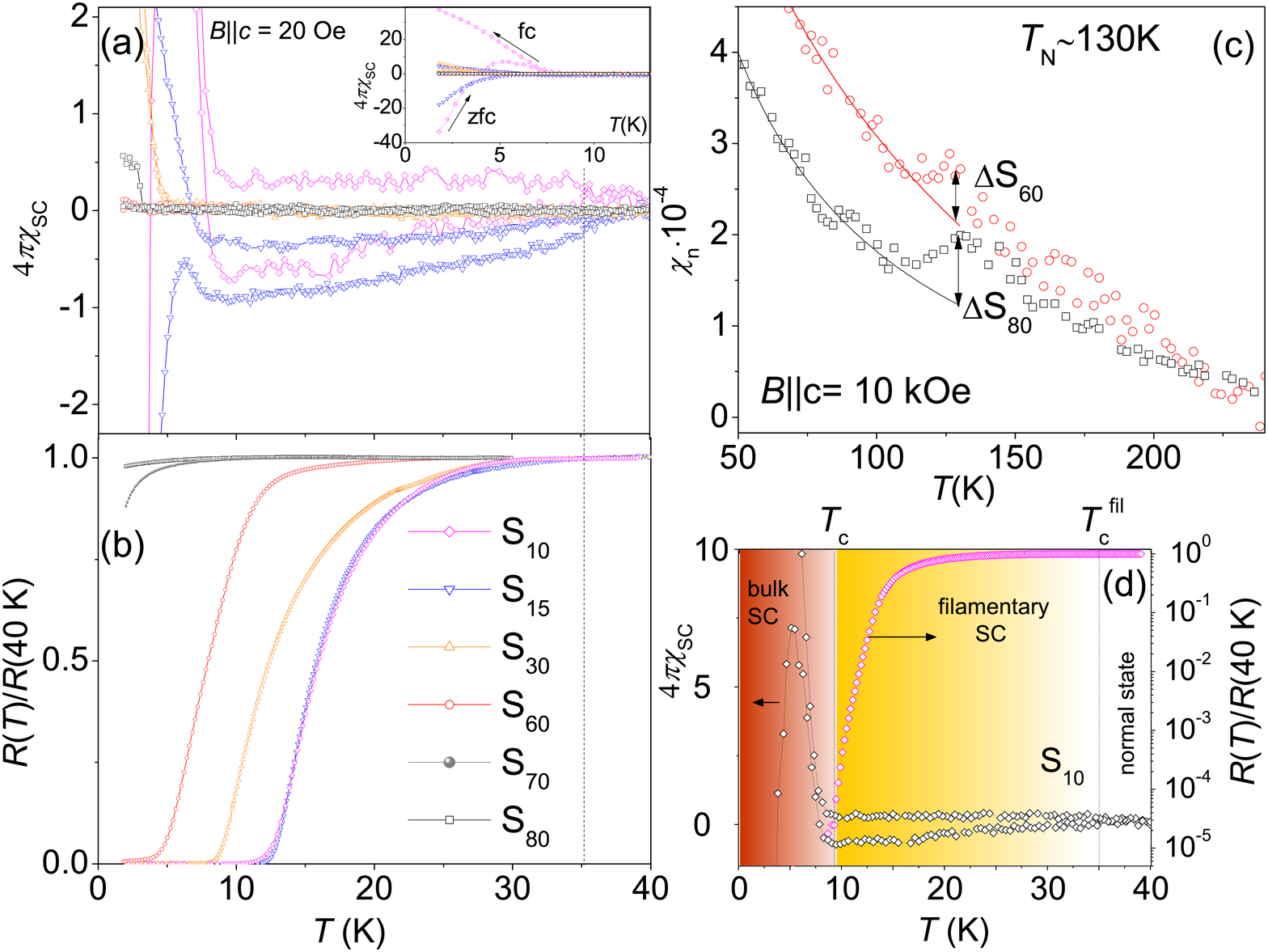}
	\caption{\textbf{Magnetic susceptibility and resistance.} 
	(a) Temperature dependence of the dimensionless magnetic susceptibility 
$\chi_{\mathrm{SC}}$(T)=($m$($T$)-$m$(40\,K))/$V_{\mathrm{Ba122}}B$ in the superconducting state of the 
films with different thickness, where $m(T)$ is sample magnetic moment, $V_{\mathrm{Ba122}}$ is the volume of the Ba122 layer and $B$ is the applied magnetic field (the data are not corrected 
by the demagnetization factors). 
The inset shows the region of bulk superconductivity.  
 (b) Temperature dependence of the normalized resistance $R$($T$)/$R$(40\,K) of the investigated thin films in zero magnetic field. 
(c) The normal state susceptibility $\chi_{\mathrm{n}}(T)=(m(T)$-$m$(300\,K))/$V_{\mathrm{Ba122}}B$. The amplitude of the SDW anomaly at $T_{\mathrm{N}}\,\approx$\,130\,K is 
reduced in the thinner film (S$_{60}$), which is partially strained. (d) Susceptibility and resistance for sample S$_{10}$ indicating two superconducting regions (bulk 
superconductivity (SC) and filamentary SC).}
	\label{figure2}
\end{figure}

The mechanism of stress relaxation is also seen in 
bright field micrographs of sample S$_{10}$ and S$_{60}$ taken by a transmission electron microscope (TEM). 
The images for the beam direction parallel to the [0\,1\,0] direction of the spinel substrate 
are shown in figs. \ref{figure1}c, d. The thickness of Fe and Ba122 layer for S$_{10}$ are $\approx$\,27\,nm and 10\,nm, respectively. 
The corresponding values for S$_{60}$ are 31\,nm and 64\,nm, respectively. 
In contrast to S$_{10}$, the sample S$_{60}$ contains cracks aligned perpendicular to the film surface of Ba122 in most cases.
Average distances between such cracks are in the range of a few $\mu$m and were also confirmed by atomic force microscopy (see Supplementary Figure S2). Also the root mean square
roughness, $R_{\mathrm{rms}}$, was determined to 3.5 nm for sample S$_{80}$. This is much 
larger in comparison to the thinner samples ($R_{\mathrm{rms}S_{30}}$\,=\,0.7 nm, $R_{\mathrm{rms}S_{10}}$\,=\,0.4 nm). 
The observed cracks are the result of the relaxation process (shortening of the 
$a$-axis of the Ba122) when the critical thickness of the Ba122 layer is 
reached. In the vicinity of the cracks the Ba122 layer is partially flaked from 
the Fe buffer layer. This leads to a formation of planar defects (probably 
several layers of BaO) parallel to the $ab$-plane as can be seen in 
fig. \ref{figure1}d.
For the films with $d\,>\,d_{\mathrm{c}}$ the observed relaxation process 
results in areas with properties similar to strain-free 
bulk Ba122 and areas which are still strained by the underlying layers (see also 
the sketches in fig. \ref{figure2}b). 
Further analysis by convergent beam electron diffraction \cite{Chekhonin2013} 
indicates that the strained state in the Ba122 layer is primarily determined by the Fe layer beneath. This is in accord with the previous observation of interfacial bonding between Fe and the FeAs sublattice of the Ba122. \cite{Thersleff2010}

\subsection*{Magnetic properties}
The SC properties of the films were investigated using several 
complementary techniques. The temperature dependence of 
the dimensionless volume magnetic susceptibility $\chi_{\mathrm{SC}}$($T$)=($m$($T$)-$m$(40\,K))/$V_{\mathrm{Ba122}}B$
and the normalized resistance $R(T)$/$R$(40\,K) for films 
with various thickness are shown in fig. \ref{figure2}, where $m(T)$ is the sample magnetic moment, $V_{\mathrm{Ba122}}$ is the volume of the Ba122 layer and $B$ is the applied magnetic field.
A decrease of $d$ results in an increase of the superconducting critical temperature and the SC volume fraction of the films. 
The upturn of the susceptibility in the SC state is probably caused by the paramagnetic Meissner effect (Wohlleben Effect) 
\cite{Luzhbin2004,Dias2004,Torre2006,Xing2009}.
The large value of the diamagnetic signal at low temperatures suggests 
that an essential volume of the films is in the SC state 
at low $T$ (see inset of fig. \ref{figure2}a). The temperature dependence of the normal volume magnetic susceptibility 
$\chi_{\mathrm{n}}(T)=(m(T)$-$m$(300\,K))/$V_{\mathrm{Ba122}}B$ for films S$_{80}$ and S$_{60}$ is shown in fig. \ref{figure2}c. 
As can be seen, the amplitude of the SDW anomaly at $T_{\mathrm{N}}\,\sim$\,130\,K is 
reduced in the thinner film (S$_{60}$), which is partially strained. Moreover,  
no signature of the SDW transition was observed for the films with $d\,<$\,60\,nm, suggesting that in-plane tensile strain effectively suppresses long-range SDW 
order, and SC does not coexist with SDW order in the thin films. $T_{\mathrm{N}}\,\sim$\,130\,K value of the unstrained thick films is consistent with $T_{\mathrm{N}}\,\approx$\,140\,K observed in stoichiometric BaFe$_2$As$_2$ single crystals \cite{Rullier-Albenque2010}. 

\subsection*{Electrical transport properties}
The SC transition width, $\Delta T_{\mathrm{c}}$\,=\,$T_{\mathrm{c,onset}}-T_{\mathrm{c}}$, 
of the films is very large (fig. \ref{figure2}b, d). For films with $d$\,$<$\,30\,nm the decrease of 
resistance starts at relatively high temperatures $T_{\mathrm{c,onset}}$\,=\,$T_{\mathrm{c}}^{\mathrm{fil}}\,\approx$\,35\,K, 
whereas zero resistance is measured at $T_{\mathrm{c}}$\,$\approx$\,10\,K as 
shown in fig. \ref{figure2}c, d. As can be seen in fig. \ref{figure2}d 
$T_{\mathrm{c}}^{\mathrm{fil}}$ exactly corresponds to the temperature at which 
splitting between zero field cooled (zfc) and field cooled (fc) branches of the susceptibility 
$\chi(T)$ curves is observed. In turn, a strong signal in 
susceptibility data for the thin films with $d$\,$<$\,30\,nm is observed below 
$T_{\mathrm{c}}$, only (fig. \ref{figure2}a). 
From the ratio between diamagnetic moments at $T\,>$\,10\,K and in 
the low-$T$ region we concluded that in the range between 
$T_{\mathrm{c}}^{\mathrm{fil}}$ and $T_{\mathrm{c}}\,\approx$\,10\,K SC exists 
only in a small volume 
fraction of the films. This so-called filamentary SC~\cite{Yamazaki2010,Colombier2009} (see Supplementary Figure S3) is 
attributed to minor regions with slightly different strain state. 
The values of $T_{\mathrm{c}}$ and $T_{\mathrm{c}}^{\mathrm{fil}}$, given in table \ref{paras} and fig. \ref{figure4}, are deduced from the method presented in fig. \ref{figure2}d.
\begin{figure}
	\centering
		\includegraphics[width=1.0\textwidth]{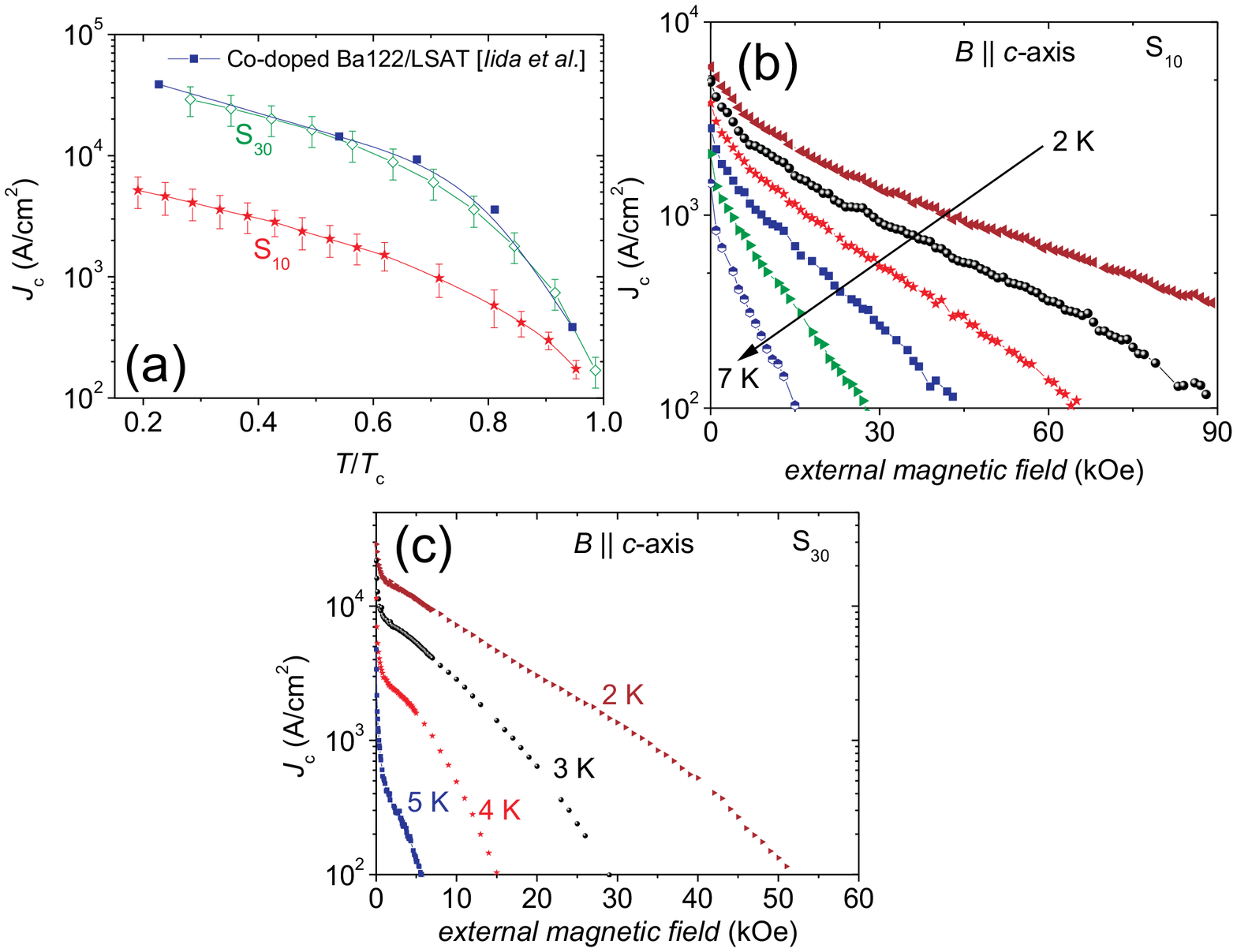}
	\caption{\textbf{Critical current densities.} 
	(a) Temperature dependence of self-field $J_{\mathrm{c}}$ for sample S$_{10}$ and S$_{30}$ compared with data taken from ref. \cite{Iida2010}. 
	The relatively large $J_{\mathrm{c}}$ values indicate bulk SC for the strained films. Bulk $T_{\mathrm{c,0}}$ was taken for normalizing the data. The error-bars are caused by the uncertainty in 
	the film thickness due to the surface roughness (minimal and maximal size of Ba122 layer thickness). (b) and (c) provide $J_{\mathrm{c}}(B)$ data for samples S$_{10}$ and S$_{30}$, respectively.}
	\label{figure3}
\end{figure}

\subsection*{Critical current densities} 
The presence of the ferromagnetic Fe buffer layer and the extremely planar geometry of the films (large demagnetization factor) do 
not allow to estimate directly the SC volume fraction from the diamagnetic screening 
at low temperatures (see fig. \ref{figure2}a). Therefore, we additionally measured the transport critical current density, $J_{\mathrm{c}}$, to demonstrate that a large volume of the films is in the 
superconducting state below $T_{\mathrm{c}}$.
It is seen in fig. \ref{figure3}a that self-field $J_{\mathrm{c}}$ values in the strained film S$_{30}$ is 
comparable with one in 
optimally Co-doped Ba122 epitaxial thin films having a similar thickness $d\approx$\,30\,nm and $T_{\mathrm{c}}\,\approx\,22$\,K.\cite{Iida2010}
We suppose that the different $J_{\mathrm{c}}$ and irreversibility field for sample S$_{30}$ and 
sample S$_{10}$ are due to different pinning properties rather than related to grain connectivity or other extrinsic factors. 
A dominant weak-link behavior is excluded by the relative robustness of $J_{\mathrm{c}}$ to external magnetic fields (see figs. \ref{figure3}b, c). Also we exclude a large variation 
of the cross section for the superconducting currents since we observed a comparable diamagnetic signal for both films (see fig. \ref{figure2}a).  
Thus, the relatively high $J_{\mathrm{c}}$ values obtained from 
transport current measurements are an additional strong evidence for a large SC 
volume of the thin films with $d\,\leq$\,30\,nm. 
   
\section*{Discussion}

\begin{figure}
	\centering
		\includegraphics[width=1.00\textwidth]{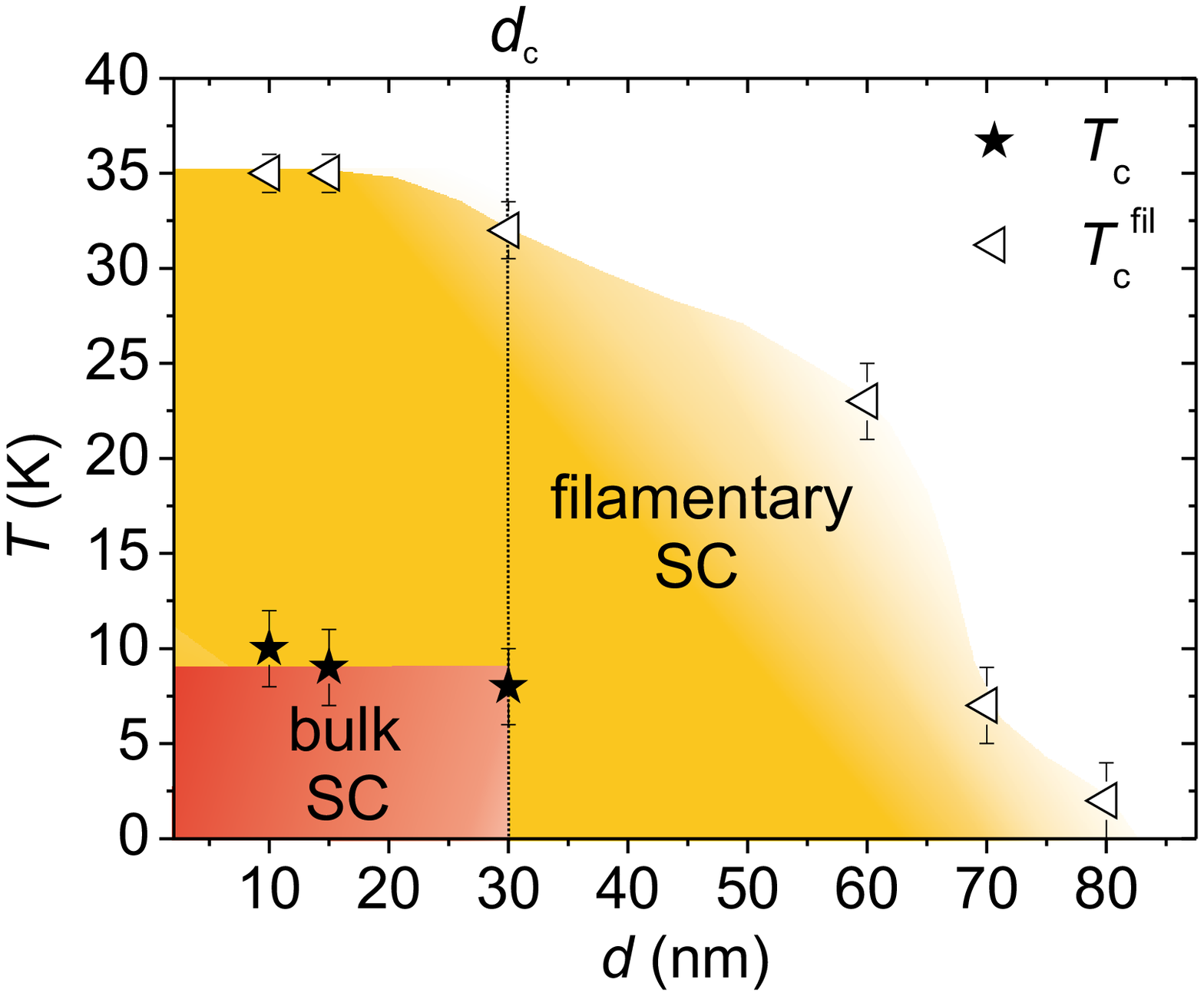}
	\caption{ 
	\textbf{Schematic $T_{\mathrm{c}}-d$ diagram for parent compound BaFe$_2$As$_2$ thin films.} 
	Filled symbols denote the critical temperature, $T_{\mathrm{c}}$, of the main superconducting phase. 
	Open symbols show the onset critical temperature $T_{\mathrm{c}}^{\mathrm{fil}}$ of the filamentary superconductivity (SC). 
	For $d_{\mathrm{c}}\,>$\,30\,nm the spin density wave (SDW) transition was observed at $T_{\mathrm{N}}\sim$\,130\,K. The colours are guide for the eyes. The error-bars were estimated by a 
	standard deviation of magnetic and transport measurement data.}
	\label{figure4}
\end{figure}

We suppose that the main mechanism of inducing SC and therefore controlling the phase diagram of Ba122 in our films is tensile in-plane strain. 
In order to investigate the possible role of charge transfer from the Fe buffer to the Ba122 layer 
similar to the case of FeSe single-layers (see introduction), 
we compared the strained films with non-strained one grown on different substrates but having similar thickness of Ba122 and Fe-buffer layers. 
It was found that the unstrained films do not show any signature of a superconducting transition (see Supplementary Figure S4). Thus, a sizable effect of 
the proximity to the metallic Fe buffer layer on $T_{\mathrm{c}}$ is ruled out. 
We sum up our main findings in Figure 4 and Table 1. As can be seen, the films with thickness $d\,<\,d_{\mathrm{c}}\,\approx$\,30\,nm having a large SC volume fraction 
are completely strained. The Ba122 layer relaxes and SC volume abruptly reduces when the thickness exceeds $d_{\mathrm{c}}$. 
The films with $d\,\sim\,$80\,nm have lattice parameters as unstrained bulk Ba122 and a negligibly small SC volume fraction. 
In order to gain a deeper understanding of the relevant parameters for tuning the Ba122 phase diagram, 
we compare our results with data of isovalently Ru-doped 
Ba122.\cite{Rullier-Albenque2010, Sharma2010} 
As can be seen in table \ref{paras} the lattice 
parameters and $T_{\mathrm{c}}$ values of overdoped Ba(Fe$_{1-x}$Ru$_x$)$_2$As$_2$ with $x\,\sim\,0.4-0.5$  
are comparable to the values of our thin films. These striking similarities evidence that 
structural changes induced by Ru doping are the dominant factor in controlling 
the phase diagram. Other factors which were assumed to explain the emergence of SC in Ru-doped Ba122 (disorder effects, dilution of the 
magnetic Fe sublattice and an enhancement in the bandwidth due to hybridization 
involving Ru \cite{Zhang2009,Dhaka2011}), only play a secondary role. From this 
point of view, disorder may explain the variations of $T_{\mathrm{c}}$ and the shifting 
of the SC dome in the phase diagram for different samples of Ru-doped Ba122.\cite{Rullier-Albenque2010, Sharma2010, Thaler2010} 
Moreover, the comparison between our data on strained Ba122 and the phase diagram of Ru-doped 
single crystals together with the observed mechanism of stress relaxation 
(as discussed above) shed a light on the filamentary SC with $T_{\mathrm{c}}^{\mathrm{fil}}\,\approx$\,35\,K found in the thinnest 
films (see fig. \ref{figure4}). The SC regions 
with $T_{\mathrm{c}}^{\mathrm{fil}}$ can be related to the minor regions with reduced strain (most 
probably due to low angle grain boundaries) of the Ba122 layer. Therefore, these regions have slightly smaller $a$ and larger $c$ lattice parameters compared to the 
main strained phase which is consistent with an analysis using electron backscatter diffraction \cite{paul2}. 
In particular, this shows that $T_{\mathrm{c}}\,\geq$\,35\,K is possible by adjusting the 
lattice strain of Ba122.    
  
Finally we conclude that the coherent interfacial bonding between Fe and the FeAs sublattice of the Ba122 results in an in-plane straining of Ba122 to the 
(001) plane of bcc Fe buffer layer, i.e. increasing $a$ and reducing $c$ lattice parameters 
of the Ba122 layer for thicknesses $d\,<\,d_{\mathrm{c}}\,\approx$\,30\,nm. 
This suppresses magnetism 
and gives rise to bulk SC with 
$T_{\mathrm{c}}\,\approx$\,10\,K. The large superconducting volume fraction of the films with 
$d\,\leq\,d_{\mathrm{c}}$ is evidenced by a pronounced diamagnetic screening and by measurements of a finite 
transport critical current. The close similarities of crystal parameters and $T_{\mathrm{c}}$ between thin films and Ru-doped Ba122 indicates that 
structural changes are one of the dominant factors for controlling $T_{\mathrm{c}}$ of Ba122. Furthermore, the observation of filamentary SC at 35\,K in regions with a different strain state 
suggests that $T_{\mathrm{c}}\,\gtrsim$\,35\,K is possible in Ba122 
without doping or applying external pressure.

\section*{Methods}

\subsection*{Sample preparation}
The targets used for the thin film growth were prepared by a conventional solid state reaction process as described in Refs.\citenum{Iida2009,kurth2012,kurth2013}. 

The thin films were grown using pulsed laser deposition with a KrF excimer laser ($\lambda$\,=\,248\,nm). The preparation process took place in an ultra-high vacuum chamber with a base pressure of 
10$^{-9}$\,mbar. Prior to the deposition the substrate was heated up to 750\,$^{\circ}$C to clean the surface. 
Subsequently, the Fe buffer layer of about 30\,nm thickness was deposited at room temperature with a laser frequency of 5\,Hz and then heated to 670\,$^{\circ}$C to 
flatten the Fe surface.\cite{Thersleff2010,Engelmann12} This temperature was held for 30 minutes before the Ba122 layer was grown with a frequency of 10\,Hz. The layer thickness was adjusted 
via the pulse number at constant laser energy and was confirmed by TEM for selected samples. 
To achieve homogeneous samples without thickness gradient the substrate was 
rotated during the whole deposition. 

\subsection*{Structural characterization}
The $c$ lattice parameters were calculated from X-ray diffractograms (in a Bruker D8 Advance Diffraktometer) 
in Bragg Brentano geometry using the Nelson Riley function, 
wheras the $a$ lattice parameters were taken out of the RSM's measured in a 
Panalytical X'pert Pro system. 

TEM characterization of the samples was performed at FEI Tecnai-T20 TEM operating at an accelerating voltage of 200\,kV. 
Preparation of the TEM lamellae was done by the focused ion beam (FIB) technique (FEI Helios 600i) using a platinum protection layer and 3\,kV accelerating voltage in the last FIB preparation step. 

Electron diffraction X-ray spectroscopy (EDX) in TEM and 
auger electron spectroscopy (AES) were performed to check the Ba122 stoichiometry. 
Sensitivity factors for the AES spectra were evaluated using single crystals.\cite{Aswartham2011} Parts of the thin film and the single crystal were sputtered, and the resulting 
spectra were analyzed in a JEOL JAMP 9500F Field Emission Auger Microprobe (see Supplementary Figure S5). 
The resulting data are given in supplement and confirm that the thin films are stoichiometric. 

\subsection*{Transport and magnetic measurements}
Transport measurements were performed using a physical property measurement system (Quantum Design) with four probe method.  
Magnetic measurements were performed in a Quantum Design DC superconducting quantum interference device (SQUID). 
In the normal state electrical currents flow 
mainly through the metallic Fe buffer layer having a lower resistance than the Ba122 layer. Hence, it is not possible to measure the SDW/structural transition via 
transport measurements. 
For the measurements of the critical current density, $J_{\mathrm{c}}$, the samples S$_{10}$ and S$_{30}$ 
were structured using ion beam etching with a stainless steel mask. 
Bridges of 0.45\,mm width and 0.6\,mm length were fabricated. $J_{\mathrm{c}}$ was determined from $E(J)$ characteristics 
using the geometry parameters of the bridges
and applying a 1\,$\mu$Vcm$^{-1}$ criterion ($J \perp B$ - max. Lorentz force configuration).

\section*{acknowledgement}

The authors would like to thank Marko Langer, Michael K\"uhnel, Steffi Kaschube and Juliane Scheiter for technical assistance and S. L. Drechsler for fruitful discussions. 
J. Engelmann and P. Chekhonin are grateful to the GRK1621 of the DFG for financial support. Additionally we thank for 
financial support from the EU (Iron-Sea under project no. FP7-283141, SUPERIRON, Grant No. 283204).
We also want to thank A. Saicharan for providing single crystals for the AES measurements. 
\\\\
\textbf{Competing financial interests}
The authors declare that they have no competing financial interests.
\\\\
\textbf{Contributions:} J.E. fabricated samples, performed $R(T)$-, AFM-, $\theta$ - 2$\theta$-, $J_{\mathrm{c}}$ measurements, analyzed the data, designed the experiments and prepared the manuscript. 
V.G. did magnetic characterisation, analyzed the data, designed the experiments and prepared manuscript. 
P.C. carried out the TEM characterisation, M.H. the AES measurements and R.H. the RSM's. F.K. and K.I. fabricated Ba122 pulsed laser deposition targets for thin-film deposition and helped 
with preparation of the thin films. J.H. and D.V.E. prepared manuscript. B.H., S.O. and W.S. supervised the experiments and contributed to manuscript preparation. 
L.S. designed and directed the research.
All authors discussed the results and implications and commented on the manuscript at all stages. 
\FloatBarrier

\begin{table*}
	\centering
	\caption{Lattice parameters of the thin films, the target, and for Ru-doped Ba122. The critical temperatures are evaluated from the $R(T)$ and the 
	$\chi(T)$ data. The Ba122-target is a polycrytalline sample. The reference samples [\citenum{Rullier-Albenque2010}] are single crystals.}
		\begin{tabular}{l c c c c c c c } \hline 
			sample & $c$-axis parameter & $a$-axis parameter  &  $c/a$ &  $V$\,=\,$a^2c$   &$T_{\mathrm{c}}^{\mathrm{fil}}$ & $T_{\mathrm{c}}$ &thickness $d$ \\ 
			 name & (nm)  &(nm) & &(\AA$^3$) & (K) &(K) & (nm)\\ \hline 
			Ba122-target \cite{kurth2012} &  1.30141(1) & 0.396 & 3.29 & 204(1) \\ 
			S$_{80}$ & 1.302(1) & 0.396(2) & 3.29 & 204(2)    & 2 & 0&$\approx$80 \\ 
			S$_{70}$ & 1.295(1) & - & - & -    & 7 & 0&$\approx$80 \\ 
			\multirow{2}{*}{S$_{60}$}  & 1.301(1) & 0.396(5) & 3.29 & 204(4)  & \multirow{2}{*}{24} & \multirow{2}{*}{0} &\multirow{2}{*}{$\approx$60}  \\ 
			                           & 1.279(1) & 0.403(5) & 3.23 & 208(4)  &                      \\
			S$_{30}$ & 1.274(1) & 0.403(2) & 3.17 & 206(2)  &32   & 8 & $\approx$30 \\ 
			S$_{15}$ & 1.271(1) & 0.404(2) & 3.15 & 207(2)  &35& 9&$\approx$15 \\
		  S$_{10}$ & 1.267(1) & 0.404(2) & 3.14 & 206(2) & 35& 10 &$\approx$10 \\
 BaFe$_2$As$_2$ \cite{Rullier-Albenque2010} & 1.3022(2) & 0.39633(4) & 3.29 & 204.55(4) & - &0&  \\ 
		  Ba(Fe$_{0.62}$Ru$_{0.38}$)$_2$As$_2$ \cite{Rullier-Albenque2010} & 1.2749(2) & 0.40342(5) & 3.16 & 207.49(5) & - &$\sim 10$&  \\ 
			\end{tabular}
		\label{paras}
\end{table*} 

\end{document}